\documentclass[prb,twocolumn,showpacs,superscriptaddress,preprintnumbers,amssymb]{revtex4-1}
\usepackage{graphicx}
\usepackage{color}
\usepackage{xcolor}
\usepackage{hyperref}
\usepackage{dcolumn}
\usepackage{bm}
\usepackage[nointegrals]{wasysym}
\usepackage{amsmath}
\usepackage{physics}

\newcommand{\beq}{\begin{equation}}
\newcommand{\eeq}{\end{equation}}
\newcommand{\beqn}{\begin{eqnarray}}
\newcommand{\eeqn}{\end{eqnarray}}

\newcommand{\ra}{\rightarrow}

\newcommand{\cA}{ {\cal A} }

\newcommand{\cC}{ {\cal C} }

\newcommand{\cL}{ {\cal L} }

\newcommand{\cS}{ {\cal S} }

\newcommand{\vect}[1]{{\bm{#1}}}
\newcommand{\Intt}{\mathrm{int}}
\newcommand{\ii}{\mathrm{i}}

\newcommand{\hrho}{\hat{\rho}}

\newcommand{\hO}{\hat{O}}

\newcommand{\U}{\mathrm{U}}

\newcommand{\cx}[1]{{\color{black} #1}}
\newcommand{\cxb}[1]{{\color{black} #1}}

\newcommand{\cmj}[1]{{\color{black} #1}}

\begin{document}

\title{ Higher-form Symmetries under Weak Measurement}

\author{Kaixiang Su}
\affiliation{Department of Physics, University of California,
Santa Barbara, CA 93106}

\author{Nayan Myerson-Jain}
\affiliation{Department of Physics, University of California,
Santa Barbara, CA 93106}

\author{Chong Wang}
\affiliation{Perimeter Institute for Theoretical Physics, Waterloo, Ontario, Canada N2L 2Y5}

\author{Chao-Ming Jian}
\affiliation{Department of Physics, Cornell University, Ithaca, New York 14853}

\author{Cenke Xu}
\affiliation{Department of Physics, University of California,
Santa Barbara, CA 93106}


\begin{abstract}

We aim to address the following question: if we start with a quantum state with a spontaneously broken higher-form symmetry, what is the fate of the system under weak local quantum measurements? We demonstrate that under certain conditions, a phase transition can be driven by weak measurements, which suppresses the spontaneous breaking of the 1-form symmetry and weakens the 1-form symmetry charge fluctuation. We analyze the nature of the transitions employing the tool of duality, and we demonstrate that some of the transitions driven by weak measurement enjoy a line of fixed points with self-duality.

\end{abstract}

\maketitle

\section{Introduction}

Quantum measurement lies at the heart of the fundamental interpretation of quantum mechanics. Measurement is one type of nonunitary process in contrast to the ordinary unitary evolution generated by a Hamiltonian. It can fundamentally change the quantum entanglement of the system. In the last few years, measurement-induced entanglement phase transition~\cite{nahummeasure,fishermeasure,PurificationTransion2020,ChoiBaoQiEhud2020,jianmeasure,BaoChoiEhud2020,GullansHuseProbes,SangHsieh2021,barkeshlimeasure,Ippoliti2021,vijaymeasure,reviewmeasure,googlemeasure,MIPTExpSC,MIPTTrapIon}, measurement prepared long-range entangled states~\cite{nat1,nat3,natnishimori,natnishimori2,leenishimori,nat4,measureTO} (topological order) have all attracted enormous interests and efforts. In particular, measurement-induced entanglement transition has been recently observed~\cite{googlemeasure,MIPTExpSC,MIPTTrapIon} in the Noisy Intermediate-Scale Quantum~\cite{nisq} platform.


Symmetry is another pillar of modern physics. The classic Landau's paradigm of understanding phases of matter is based on the notion of symmetry and spontaneous symmetry breaking (SSB). For decades ``non-Landau" physics, such as topological order and spin liquid, has been one of the most active directions of condensed matter physics. However, in the last few years, the generalized symmetries such as higher-form symmetries~\cite{formsym0,formsym4,formsym6,formsym7,formsym8,formanomaly,mcgreevyreview} have made Landau's paradigm much more inclusive and many non-Landau physics can be unified in a generalized Landau's paradigm. In particular, a topological order can often be interpreted as the SSB phase of a higher-form discrete symmetry, and the emergent photon phase~\cite{photonhermele,photonsondhi,photonwen} of (for example) the quantum spin ice~\cite{spinicemonopole,spinicebenton} can be interpreted as the SSB phase of an (emergent) $\U(1)$ 1-form symmetry.

A higher-form symmetry can have a very different phenomenology from an ordinary 0-form symmetry. For example, the ground state degeneracy arising from SSB of a 1-form symmetry (which used to be interpreted as topological degeneracy) is robust against explicitly breaking the 1-form discrete symmetry; the photon excitations which correspond to the Goldstone modes~\cite{goldstone1,goldstone2} arising from spontaneous breaking of the $\U(1)$ 1-form symmetry is also robust against the weak explicit breaking of the 1-form symmetry~\cite{mcgreevy2,wen2023}; \cx{the Goldstone modes can only be gapped with strong enough breaking of the 1-form symmetry, after a ``Higgs transition"}. All these are very different from the phenomenology of 0-form symmetries. For example, when we explicitly break a 0-form symmetry, the Goldstone mode would immediately acquire a mass gap and become a pseudo-Goldstone mode.

In this work, we will show that higher symmetry and 0-form symmetry also have very different behaviors under weak quantum measurement. If we start with an SSB phase of a 0-form symmetry, a weak measurement cannot drive an intrinsic phase transition; while an SSB phase of a higher-form symmetry can go through a phase transition driven by weak measurement. These transitions have a natural description on a temporal defect of the Euclidean space-time. We will demonstrate that weak-measurement-driven transition for $\U(1)$ 1-form symmetry may possess a line of fixed points connected through self-duality; In an upcoming work, we will show that, under weak measurements, the properties of quantum states with discrete 1-form symmetries may be mapped to those in a series of classical statistical mechanics models, some of which have well-known phase transitions.

\section{General formula}

Without loss of generality, a pure density matrix $\hrho_0$, after going through a series of local weak measurements, 
are transformed according to
$\hrho_0 \rightarrow \hrho = \mathcal{E}[\hrho_0]$, where $ \mathcal{E}[\hrho_0]$ is a composition of local measurements $\mathcal{E}[\hrho_0] = \otimes_{\vect{x}} \mathcal{E}_{\vect{x}}[\hrho_0]$ where \beqn \mathcal{E}_{\vect{x}}[\hrho_0] = (1 - p) \hrho_0 + p \left( \sum_m p_m \hat{P}_{m,\vect{x}} \hrho_0 \hat{P}_{m,\vect{x}} \right). \eeqn 
\cx{In this formula, $p$ measures the strength of the system being entangled with the ancilla qubits (please refer to the suplementary material for an explicit example)}; $\hat{P}_{m,\vect{x}}$ is a projection operator that projects the system to the measurement outcome $m$, the measurement outcomes are summed up with weight $p_m$. \cx{When there is no ``post-selection" of the measurement outcomes, $p_m = 1$, the mapping $\cal E$ is a quantum channel in this case. With post selections, $p_m$ can be more general. This work considers the effect of weak measurements with and without post selections.} ~\footnote{When $p_m$ differs from 1, $\cal E$ generically yields un-normalized density matrices that must be normalized. The resulting transformation on $\hat \rho_0$ becomes $\hat \rho_0 \rightarrow \mathcal{E}[\hrho_0]/\tr{\mathcal{E}[\hrho_0]}$, which is a non-linear map of the density matrix. }

\begin{center}
\begin{figure}
\includegraphics[width=0.4\textwidth]{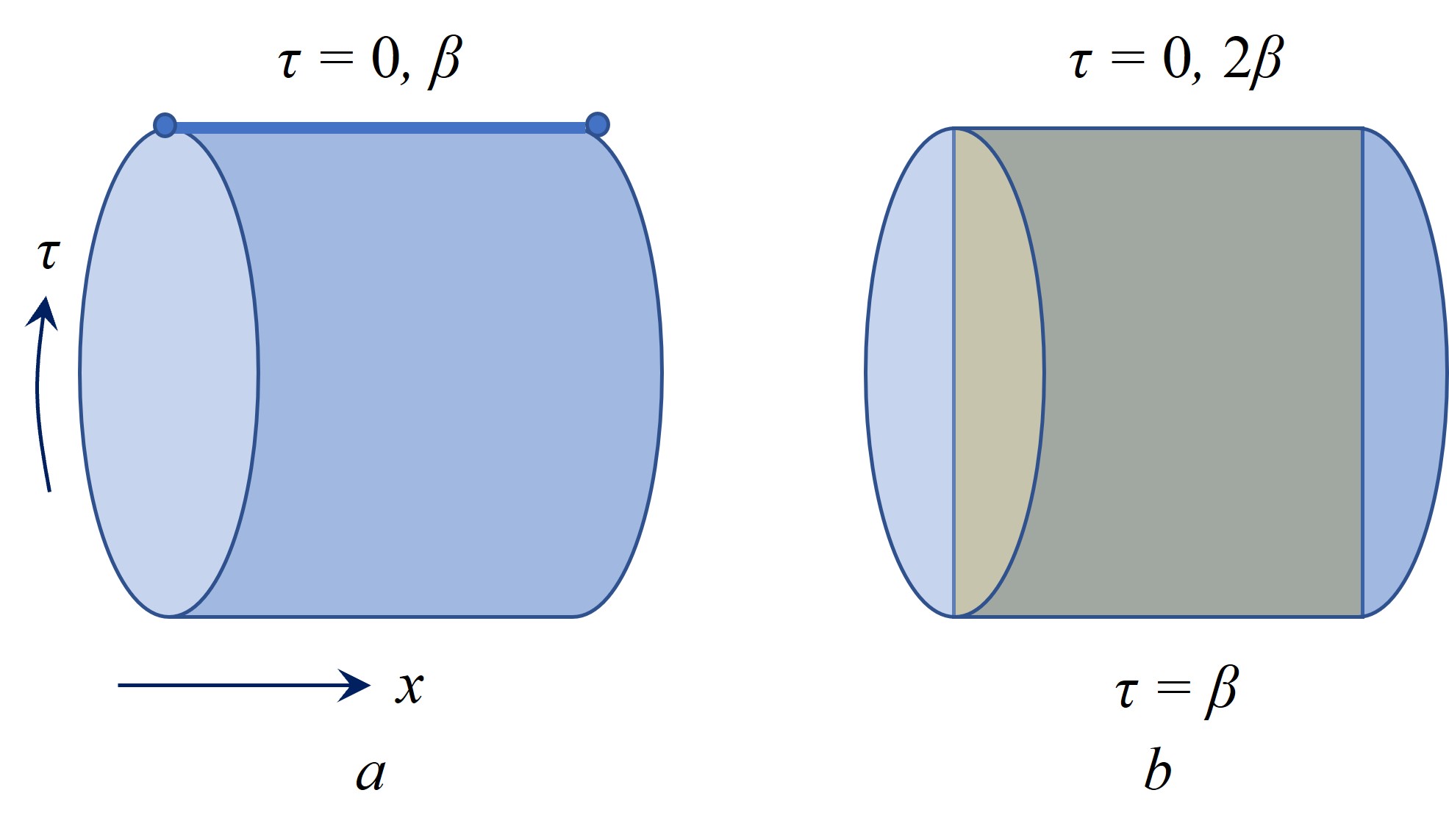}
\caption{The Euclidean space-time picture of quantities $\tr\{\hrho \ \hat{O}\}$ ($a$), and the 2nd R\'{e}nyi entropy
($b$). } \label{spacetime}
\end{figure}
\end{center}

For a generic complicated quantum many-body system, deriving the exact microscopic density matrix is difficult and often unnecessary. The Euclidean space-time path integral formalism allows us to use various analytical techniques, such as coarse-graining and the renormalization group (RG), to capture the physics in the infrared limit. In this formalism, a density matrix $\hrho_0$ of a pure state $|\Psi\rangle$ is evaluated as $[\hrho_0]_{\phi_1(\vect{x}), \phi_2(\vect{x})} \sim \lim_{\beta \ra \infty } \int D \phi(\vect{x}, \tau) \exp \left( - \cS \right) $, with the boundary condition $ \phi(\vect{x},0) = \phi_1(\vect{x})$ and $\phi(\vect{x},\beta) = \phi_2(\vect{x})$. Here $\cS$ is the action of the system whose ground state is the desired pure state $|\Psi\rangle$. 
After the weak measurements (with or without post-selection),
the density matrix becomes~\cite{wfdecohere}
\beqn && [\hrho]_{\phi_1(\vect{x}), \phi_2(\vect{x})} \sim \lim_{\beta \ra \infty } \int \ D \phi (\vect{x}, \tau) \exp\left( - \cS - \cS^{\Intt} \right); \cr\cr && \cS^{\Intt} = \int d\vect{x} \ \cL^{\Intt} ( \phi(\vect{x}, 0), \phi(\vect{x},\beta)). \label{mix} \eeqn
In this path integral, the aforementioned boundary condition $ \phi(\vect{x},0) = \phi_1(\vect{x})$ and $\phi(\vect{x},\beta) = \phi_2(\vect{x})$ still holds. Hence the effect of weak measurements is represented by extra terms on the two temporal boundaries $\tau = 0$ and $\tau = \beta$, including the interaction between the two temporal boundaries. The form of $\cL^{\Intt}$ can be determined by the symmetry condition of the density-matrix transformation $\cal E$ induced by measurements, either a ``doubled" (or strong) symmetry condition or a ``diagonal" (or weak) symmetry condition~\cite{groot,sptdecohere}. Suppose we always look at quantities linear with the density matrix such as $\tr\{ \hrho \hat{O} \}$, the temporal boundaries $\tau = 0$ and $\tau = \beta$ are ``glued" together. Our system is mapped to a problem in Euclidean space-time with a defect at the temporal slab $\tau = 0,\beta$ (Fig.\ref{spacetime}$a$). If the system has a Lorentz invariance in the infrared, after the space-time rotation, the temporal defect will become the physical spatial defect in space-time~\cite{altman1,wfdecohere}.

In this work, we consider systems with an $n$-form symmetry $G^{(n)}$. We will always use $d$ to label the spatial dimension and $D$ to label the space-time dimension. There are several different questions that one can ask. The first potential question is that if we start with a $d-$dimensional state $|\Psi\rangle$ that is a gapped disordered state and also symmetric under symmetry $G^{(n)}$, what can happen to the system under weak measurements that preserve $G^{(n)}$? In our space-time picture, this question is equivalent to asking what happens to the boundary or defect of the system with $d-1$ spatial dimensions under the extra defect Hamiltonian. Since the bulk is disordered and gapped, we can often integrate out the bulk to renormalize the defect Hamiltonian. The IR physics of our problem is often identical to a local $(d-1)$-dimensional system. Some examples of this sort were discussed in Ref.~\onlinecite{wfdecohere,altman2,fan2023}. If we start with a $2d$ Ising disordered phase $|\Psi\rangle$, weak measurement can drive a phase transition which belongs to the $(2+0)d$ Ising or equivalently $(1+1)D$ Ising universality class, rather than a $(2+1)D$ Ising universality class which corresponds to an equilibrium quantum phase transition for a $2d$ quantum Ising system. This transition is also dual to an information transition of the toric code topological order under dephasing~\cite{kitaevpreskill}.

Hence in this paper, we focus on another class of questions that are of potentially greater interest: we consider a system with a symmetry $G^{(n)}$, but we start with a state $|\Psi\rangle$ with spontaneous symmetry breaking (SSB) of $G^{(n)}$, then what can happen to the system under weak measurements? Though the boundary-defect mapping is still valid, in this case, the bulk could have gapless Goldstone modes arising from the SSB and interfere with the defect physics in a much more nontrivial way.

If our system has only a 0-form symmetry, the answer to the question above is straightforward: A spontaneous breaking of a 0-form symmetry with true long-range order {\it cannot} be qualitatively altered by any local finite-strength weak measurement with or without post-selection; the SSB can only be destroyed with ``strong" projective measurements, i.e. $p=1$. The physical picture of this conclusion is that there is a strong ``proximity effect" from the bulk, which breaks the 0-form symmetry on any defect with one lower dimension. More formally, for a true long-range order that spontaneously breaks a 0-form symmetry, the state can be exemplified as a simple product state. For example, for an Ising system whose global $\mathbb{Z}^{(0)}_2$ charge is $\prod_j X_j$, an Ising-ordered state has the following wave function at its fixed point: $ |\Psi\rangle = \prod_{j} | Z_{j}  = 1 \rangle$. Now on every site, we perform a weak measurement on operator $X_{j}$ with probability $p$, namely, there is a probability $p$ that the qubit on the site $j$ is being measured for each $j$. The density matrix takes the following form when the measurement outcome $X_{j} = 1$ is selected \beqn \rho = \otimes_{j} \rho_j, \ \ \ \rho_j = (1 - p) \frac{I + Z_{j}}{2} + p \frac{I + X_{j}}{2}. \eeqn Then as long as $p < 1$, the expectation value of $Z_j$ is always nonzero. And there is no intrinsic phase transition driven by the weak measurements as the system essentially reduces to a 0-dimensional problem. This conclusion holds {\it with or without } post-selecting the measurement outcomes.

We will show that higher-form symmetries lead to much more interesting consequences, as the SSB phase of the higher-form symmetry is {\it not} a simple product state in space, and an intrinsic phase transition can be driven by weak measurements.

\section{Higher-form symmetries}

This section starts with SSB phases of $\U(1)$ higher-form symmetries. Let us first list our main conclusions:

{\bf 1.} SSB of a $\U(1)$ 1-form symmetry ($\U(1)^{(1)}$-SSB) in spatial dimension $d \geq 3$ (space-time dimension $D \geq 4$) can be qualitatively suppressed by finite strength weak measurements through a phase transition. In particular, at $d = 3$ the phase transition driven by weak measurements at the critical strength has a line of fixed points with self-duality. Here, the suppression of SSB is signified by the qualitatively weakened fluctuation/correlation of the 1-form symmetry charges. 

{\bf 2.} $\U(1)^{(2)}$-SSB can be suppressed in spatial dimension $d \geq 4$ through a phase transition driven by weak measurements. A similar pattern continues to higher-form symmetries at higher dimensions.

For lower dimensional systems, the fate of $\U(1)^{(n)}$-SSB phases under weak measurements can be perceived through duality. A $\U(1)^{(1)}$-SSB phase in $2d$, and a $\U(1)^{(2)}$-SSB phase in $3d$ are both dual to a superfluid phase, i.e. they are dual to phases with SSB of a 0-form symmetry. And as we argued in the previous section, a generic SSB phase of a 0-form symmetry cannot be qualitatively altered by finite-strength weak measurements.

\subsection{$\U(1)^{(1)}$ 1-form symmetry in $3d$}

In this section, we consider a $3d$ system with a $\U(1)^{(1)}_m$ symmetry, and this $\U(1)^{(1)}_m$ is spontaneously broken in state $|\Psi\rangle$. In the traditional language, the state $|\Psi\rangle$ has gapless photon excitations described by a gauge field $a_\mu$. A $\U(1)^{(1)}_m$ symmetry implies that there is no dynamical Dirac monopole in the system. We also assume that the electric charges that couple to $a_\mu$ are bosonic and gapped. Hence, below the gap of the electric charges, there is also an emergent $\U(1)^{(1)}_e$ symmetry in the infrared. Both the exact $\U(1)^{(1)}_m$ symmetry and the IR-$\U(1)^{(1)}_e$ symmetry are spontaneously broken in the state $|\Psi\rangle$.

For a quantum system at equilibrium, the quantum phase transition from a phase with $\U(1)^{(1)}_m$ and IR-$\U(1)^{(1)}_e$ SSB to the $\U(1)^{(1)}_m$ symmetric phase is a Higgs transition driven by the condensation of the electric gauge charge in the $(3+1)D$ space-time, by gradually increasing the kinetic energy of the electric charges. This transition is likely a weak first-order phase transition~\cite{weinberg}. The condensation of the electric charge in the $(3+1)D$ system will restore the $\U(1)^{(1)}_m$ symmetry, and renders the IR-$\U(1)^{(1)}_e$ symmetry absent.

\cx{In the supplementary material~\footnote{See supplementary material (URL) for an explicit measurement protocol driving a transition described here, which includes references~\cite{abelianhiggs1,abelianhiggs2}} we will devise an explicit measurement protocol that can drive a charge condensation (at the temporal slab $\tau = 0, \beta$), for both a discrete gauge theory and a U(1) gauge theory. Here we focus on the infrared universal physics of the condensation transition of the electric charges through weak measurements.} We will first evaluate quantities linear with $\hrho$, with the form $\tr\{\hrho \ \hO \}$. 
This Higgs transition on the $(3+0)d$ defect is described by the following ``nonlocal" bosonic QED
\beqn \cS = \int d^3x \ |(\vect{\nabla} - \ii \vect{a}) \phi|^2 + r |\phi|^2 + g |\phi|^4 + \frac{1}{2e^2} \vect{f} \cdot \frac{1}{|\partial|} \vect{f}, \label{defectqed} \eeqn where $\vect{f} = d\vect{a}$. Here $\vect{a}$ is the three spatial components of $a_\mu$ along the $(3+0)d$ defect; $\phi$ describes the electric charge that couples to gauge field $\vect{a}$. The nonlocal action of $\vect{a}$ in the last term of Eq.~\ref{defectqed} arises from the fact that $a_\mu$ lives in the entire $(3+1)D$ bulk with an ordinary Maxwell action $ \cS_{\mathrm{bulk}} \sim \int d^4x f_{\mu\nu} f_{\mu\nu} /(4e^2)$. After integrating the momentum orthogonal to the defect plane, we obtain the nonlocal action of $a_\mu$ in Eq.~\ref{defectqed}~\cite{sondual}.

The reason Eq.~\ref{defectqed} is interesting is that the coupling constant $e$ of the nonlocal QED is exactly marginal, and Eq.~\ref{defectqed} enjoys a {\it self-duality}~\cite{motrunichdual,sondual,mrossdualPRL}. We can perform the standard particle-vortex duality for the $\phi$ field in $(3+0)d$ through the following duality relation~\cite{peskindual,halperindual,leedual}: $ \ast \vect{J}^{\phi} \sim d \vect{b}$,  where $\vect{J}^\phi$ is the charge current on the $(3+0)d$ defect; and $\vect{b}$ is the dual gauge field that couples to the vortex field $\phi_v$ of $\phi$. After integrating out $\vect{a}$, Eq.~\ref{defectqed} is transformed into its dual form \beqn \cS_d = \int d^3x \ |(\vect{\nabla} - \ii \vect{b}) \phi_v|^2 - r |\phi_v|^2 + \tilde{g} |\phi_v|^4 + \frac{1}{2\tilde{e}^2} \tilde{\vect{f}} \cdot \frac{1}{|\partial|} \tilde{\vect{f}}, \label{defectdual} \eeqn where $\tilde{\vect{f}} = d \vect{b}$. Notice that Eq.~\ref{defectdual} takes the same form as Eq.~\ref{defectqed}, and the dual coupling constant $\tilde{e} \sim 4\pi / e$.

A $\U(1)^{(1)}_m$-SSB phase is a condensate of the magnetic flux (the 1-form charge of the $\U(1)^{(1)}_m$ symmetry). The condensation of the electric charge ($r < 0$ in Eq.~\ref{defectqed}) would drive the magnetic flux into an uncondensed phase on the $(2+1)D$ defect. In the dual theory, the vortex $\phi_v$ carries a magnetic flux orthogonal to the defect plane. Hence when the vortex condenses ($r > 0$), it drives the magnetic flux back into the condensate. Hence the vortex condensate with $r > 0$ in Eq.~\ref{defectdual} corresponds to the original $\U(1)^{(1)}_m$-SSB phase.

The condensation of $\phi$ at the defect will lead to certain boundary conditions for the bulk gauge field $a_\mu$. At the $(3+0)d$ temporal defect $\vect{a}$ will acquire a Higgs mass in the condensate of $\phi$, which becomes relevant and enforces a Dirichlet boundary condition for $\vect{a}$, analogous to the ``ordinary boundary condition" in the context of boundary criticality, near the temporal defect. This boundary condition implies that $\vect{a}$ should be replaced by $\partial_\perp \vect{a}$ near the defect, and hence acquires a larger scaling dimension than the bulk. This also leads to the Dirichlet boundary condition for the magnetic field $\vect{B} = \vect{\nabla} \times \vect{a}$. Then if one measures the correlation of $\vect{B}$ after the phase transition driven by the weak measurements, it will scale as \beqn \tr\{ \hrho \ \vect{B}(0) \cdot \vect{B}(\vect{x}) \} \sim \frac{1}{|\vect{x}|^6}, \eeqn rather than $1/|\vect{x}|^4$ as in the bulk of the system with SSB of $\U(1)^{(1)}_m$ and IR-$\U(1)^{(1)}_e$. If $\vect{B}$ and $\vect{a}$ are periodically defined, we can compute the following quantity \beqn \log \left( \tr\{ \hrho \ e^{\ii B(0)_a} e^{- \ii B(\vect{x})_b} \} \right) \sim \frac{1}{|\vect{x}|^6}.
\eeqn Hence the condensation of charges induced by weak measurement still weakens the correlation of the magnetic field, which is the charge of $\U(1)^{(1)}_m$; but unlike the condensation of charges in equilibrium, the correlation of magnetic field is not rendered fully short-ranged. 

After a Wick-rotation, the condensate of $\phi$ in the $(3+0)d$ defect becomes a thin-film superconductor in the $2d$ space (the XY plane), or $(2+1)D$ space-time. Then it is known that the in-plane components of the electric field $(E_x, E_y)$ of the gauge field $a_\mu$ must vanish due to screening from the condensed charges, and the out-of-plane magnetic field $B_z$ is zero inside the superconductor due to the Meissner effect. These are equivalent to the Dirichlet boundary conditions for $E_x, E_y$, and $B_z$. Then after a space-time rotation in the $(z, \tau)$ plane, 
$(E_x, E_y)$ becomes $(B_y, B_x)$. Hence the boundary condition imposed by a physical thin film superconductor precisely becomes the Dirichlet boundary condition mentioned in the previous paragraph.

Since the condensate of electric charges at the $(3+0)d$ defect does not impose the Dirichlet boundary condition for the electric field, the electric field $\vect{E}$ correlation function is expected to be qualitatively similar to the original pure state of the $\U(1)^{(1)}_m$-SSB phase. The correlation of $\vect{E}$ and $\vect{B}$ under weak measurement can also be derived using the explicit wave functional of the $\U(1)^{(1)}_m$-SSB state, which will be presented in the appendix.

Another quantity of interest is the 't Hooft loop operator $\exp(\ii \oint_{\cC} d\vect{l}\cdot \vect{b})$ with $\vect{E} = \vect{\nabla} \times \vect{b}$, whose expectation value decays as a perimeter law in the $\U(1)^{(1)}_m$-SSB pure state. It turns out that the weak measurement won't qualitatively alter the perimeter law of the 't Hooft loop expectation value, even if the charges condense on the $(3+0)d$ defect in the Euclidean space-time, which is in stark contrast to the Higgs phase in the entire $(3+1)D$ bulk. The physical picture of this result is that the 't Hooft loop operator is also a membrane operator with $\oint_{\cA} d\vect{\sigma} \cdot \vect{E} $, with $\cC = \partial \cA$. When we evaluate the 't Hooft loop $\cC$ on the $(3+0)d$ defect, the most favorable membrane $\cA$ to choose is not the one that is strictly localized on the defect, but the one that ``leaks" into the bulk, which still yields a perimeter law decay rather than area law. In the appendix, we will show that the condensate of charges on the $(3+0)d$ defect driven by weak measurement leads to a larger coefficient of the perimeter-law, which also means that the SSB of $\U(1)^{(1)}_m$ is suppressed. 


Quantities nonlinear with the density matrix, such as the 2nd R\'{e}nyi entropy $S^{(2)} = - \log \left( \tr \{ \hrho^2 \} \right)$ is of information-theoretic interests, and it can potentially also go through a phase transition under weak measurement, {\it with or without} post-selection. In particular, when there is a phase transition induced by weak measurement, we expect to see singularity in the second R\'{e}nyi entropy. The second R\'{e}nyi entropy amounts to performing path integral of fields in the duplicated space-time, with $\tau \in [0, 2\beta]$, and weak measurement is mapped to the nonlocal interaction between the $3d$ temporal slabs $\tau =
0$ and $\tau = \beta$ (Fig.~\ref{spacetime}$b$).

Ref.~\onlinecite{wfdecohere,altman2,fan2023} showed that when we consider the 2nd R\'{e}nyi entropy, quantum decoherence (or weak measurement without post-selection) can drive the condensation of ``paired" objects, which is a nonlocal bound state of objects on $\tau = 0$ and $\tau = \beta$ slabs in Fig.~\ref{spacetime}$b$. In the supplementary material we will also discuss a similar transition in the 2nd R\'{e}nyi entropy for systems with U(1) 1-form symmetries.



\subsection{Higher dimensions and higher-form symmetries}

In general dimension $d > 3$, if there is an exact $\U(1)^{(1)}_m$ symmetry in the $D = d+1$ dimensional bulk, the electric charge field $\phi$ is now a $(d-3)$-dimensional membrane in space, which couples to a $(d-2)$-form $\U(1)^{(d-2)}_e$ electric gauge field $a$. If we still start with a $\U(1)^{(1)}_m$-SSB phase, through measurement on the kinetic energy of $\phi$ membrane, a transition of $\phi$ condensation can be driven to Higgs the $(d-2)$-form $\U(1)^{(d-2)}_e$ gauge field $a$. Condensation of a membrane-like object can be tedious to describe, but fortunately this transition can still be described conveniently in the dual picture, through a similar dual relation: $\ast J^\phi \sim db$ on the $D = d$ dimensional defect. Here $J^\phi$ is a $(d - 2)$-form current of the $\phi$-membrane on the defect, and $b$ is {\it always} a 1-form gauge field. The 1-form gauge field $b$ is still coupled to a scalar matter field $\phi_v$. The transition we are after is still described by the same action as Eq.~\ref{defectdual}, with dimension $D = d$.
$\phi_v$ always describes a segment of 1-form charge of the original $\U(1)^{(1)}_m$ symmetry, which is a vector (analogous to the magnetic field $\vect{B}$) that orthogonally pierces through the defect. When $\phi_v$ condenses ($r > 0$), the $\phi$ membrane is suppressed; and when $\phi_v$ is gapped ($r < 0$), the $\phi$ membrane condenses.

But we can see that when $d > 3$, the coupling constant $\tilde{e}$ is {\it irrelevant} at the critical point $r = 0$, more precisely the scaling dimension of $\tilde{e}^2$ is $\Delta_{\tilde{e}^2} = 3 - d$. Hence the transition described by Eq.~\ref{defectdual} would be at the Gaussian fixed point for general spatial dimension $d > 3$.


In spatial dimension $d > 3$, if there is an exact $\U(1)_m^{(d-2)}$-form symmetry in the bulk in its SSB phase, weak measurements can drive a transition to suppress the SSB. This transition is also described by Eq.~\ref{defectqed} on the defect with $D = d$ dimensional space-time. Here $a$ is the dual 1-form gauge field of the $(d-2)$-form $\U(1)_m^{(d-2)}$ symmetry that is spontaneously broken in the bulk, and the $\phi$ field is the charge of the 1-form gauge field $a_\mu$, or the ``Dirac monopole" of the $\U(1)_m^{(d-2)}$ gauge field in spatial dimension $d$. Now the coupling constant $e$ is again irrelevant at the Gaussian fixed point of $\phi$. Hence, the transition driven by weak measurements is also ``mean-field" like.

\section{Summary and Discussions}

In this work, we discuss the fate of higher-form symmetry under weak quantum measurements. In particular, we demonstrate that a pure state with SSB of higher-form symmetry can go through a phase transition driven by weak measurements. This starkly contrasts with the SSB state of a 0-form symmetry whose fate under weak measurements reduces to a $(0+1)d$ problem and hence won't have an intrinsic phase transition. Some of the phase transitions driven by weak measurement can be mapped to a mixed-dimensional QED, with a line of fixed points mapped to each other under self-duality.

There is another ingredient of the mixed dimensional QED that plays a nontrivial role in the duality, i.e., the topological $\Theta-$term in the bulk. In our current set-up, the effect of the $\Theta-$term likely cancels out at the temporal defects due to the trace of the density matrix. But the $\Theta-$term can make nontrivial contributions to quantities such as the strange correlator~\cite{YouXu2013}, which involves temporal interfaces between two different states. It is worth exploring weak-measurement-driven transitions of strange correlators of systems with higher-form symmetries, which we expect to have an even richer structure of duality due to the nontrivial role of the $\Theta-$term. 

In this work, we focused on SSB states of {\it continuous} higher-form symmetries. In another work~\cite{su2024}, we will demonstrate that, if one starts with an SSB state at $d$-spatial dimension ($(d+1)D$ space-time) with a {\it discrete} higher-form symmetry, 
under different conditions various classical statistical mechanics models can be engineered through weak measurement.

C.X. acknowledges support from the Simons Foundation through the Simons Investigator program. C.-M.J. is supported by a faculty startup grant at Cornell University. Research at Perimeter Institute (C.W.) is supported in part by the Government of Canada through the Department of Innovation, Science and Industry Canada and by the Province of Ontario through the Ministry of Colleges and Universities. This research was supported in part by grant NSF PHY-2309135 to the Kavli Institute for Theoretical Physics (KITP).


\appendix

\onecolumngrid

\section{Weak-measurement on $Z_2$ gauge theory}

\cx{Before discussing our measurement protocol on systems with U(1) gauge fields, we first discuss a simpler example of weak-measurement driven phase transition on the $Z_2$ gauge theory. More examples of measurement-driven transitions on systems with discrete higher form symmetries (discrete gauge theories) are discussed in Ref.~\cite{su2024}.} We take the $3d$ toric code Hamiltonian as a starting point, which is defined on the $3d$ cubic lattice
\begin{equation}
    H_{\rm TC} = -\sum_{v} A_v - \sum_{p} B_p
\end{equation}
where $v$ denotes vertices on the cubic lattice and $p$ denotes plaquettes on the cubic lattice. The local Hilbert space is made of qubits that reside on the links. $A_v$ is then the product of $\sigma^x$ operators connected to the vertex $v$; $B_p$ is the product of $\sigma^z$ operators along the edge of plaquette $p$. $\sigma^x_{ij}$ is the analogue of the electric field of the $\U(1)$ gauge field, and $\sigma^z_{ij}$ is the analogue of $e^{\ii a_{ij}}$ of the lattice $\U(1)$ gauge field. 

\cxb{The initial state is prepared to be the tensor product between the ground state of $H_{\rm TC}$ and $\ket{0}_a$ of the ancilla qubits.}  We then evolve the system qubit and ancilla qubit via the following unitary operator
\begin{equation}
    U = \frac{1}{\sqrt{2}}\begin{bmatrix}
        \cos t + \sigma^z \sin t , \ -\cos t + \sigma^z \sin t \\
        \cos t - \sigma^z \sin t , \ \cos t + \sigma^z\sin t 
    \end{bmatrix}
\end{equation}
\cmj{Here, the $2\times 2$ matrix structure of $U$ is associated with the 2-fold Hilbert space of the ancillary qubit.}
When $t=0$, $U$ takes the form
\begin{equation}
        U_0 = \frac{1}{\sqrt{2}}\begin{bmatrix}
        1 & -1 \\
        1 & 1
    \end{bmatrix}
\end{equation}
The general $U(t)$ can be implemented via the following procedure
\begin{equation}
    U = e^{- \ii t H}\cdot U_0
\end{equation}
with
\begin{equation}
    H = \ii \sigma^z \sigma_a^+ - \ii \sigma^z \sigma_a^{-}
\end{equation}
where $\sigma_a$ denotes the Pauli operator acting on the ancilla qubit. We then measure the ancilla in its computational basis which gives us an ensemble of states labeled by measurement outcome $s$, a string that records the measurement outcomes on each link
\begin{equation}
    \ket{\psi(s)} \propto \prod_{\rm link}\frac{1}{\sqrt{2}}(\cos t + s_{ij} \sin t \sigma^z_{ij})\ket{\psi}
\end{equation}
To calculate expectation values of quantities of this ensemble, we need to evaluate
\begin{equation}
    \Bar{\expval{O}} = \sum_{s} P(s) \frac{\bra{\psi(s)}O \ket{\psi(s)}}{\braket{\psi(s)}{\psi(s)}}
\end{equation}
By Born's rule (no bias), this is equivalent to
\begin{equation}
    \tr \left( \prod_{\rm link} \frac{1}{\sqrt{2}}(\cos t + \sin t \sigma^z_{ij})\hrho_0 \frac{1}{\sqrt{2}}(\cos t + \sin t \sigma^z_{ij})O + \prod_{\rm link} \frac{1}{\sqrt{2}}(\cos t - \sin t \sigma^z_{ij})\hrho_0 \frac{1}{\sqrt{2}}(\cos t - \sin t \sigma^z_{ij})O \right)
\end{equation}
Assuming $O$ commute with $\sigma^z_{ij}$, the equation above is equivalent to
\begin{equation}
    \prod_{\rm link} \tr (\hrho_0 O)
\end{equation}
hence the measurements without post-selection does not change the expectation value of $O$. 

We now consider a biased selection, meaning we collect all measurement outcomes but we give a higher weight to the outcome with $s_{ij} = +1$. Now the expectation of $O$ becomes (again we consider quantity $O$ that commutes with $\sigma^z_{ij}$)
\begin{align*}
    &\tr \left( \prod_{\rm link} (\frac{1}{2}+x)(\cos t + \sin t \sigma^z_{ij})(\cos t + \sin t \sigma^z_{ij})\hrho_0 O + \prod_{\rm link} (\frac{1}{2}-x) (\cos t - \sin t \sigma^z_{ij})(\cos t - \sin t \sigma^z_{ij})\hrho_0 O \right) \\
    =&\tr \left( \prod_{\rm link} (1+ (4x \cos t \sin t) \sigma^z_{ij} )\hrho_0 O \right) \\
    =& \tr \left( \prod_{\rm link} (1+ (\tanh \tilde{\beta}) \sigma^z_{ij})\hrho_0 O \right)
\end{align*}
Here, we make the identification $4x \cos t \sin t = \tanh \tilde{\beta}$. We can reorganize the mixed density matrix into the following form:
\begin{equation}
   \hrho \sim e^{- \frac{1}{2} I\left( \{ \sigma^z_{ij} \} \right) } \hrho_0   e^{- \frac{1}{2} I\left( \{ \sigma^z_{ij} \} \right) },  \ \ \  \tr\left( \hrho O \right) \sim \tr \left( \hrho_0 e^{- I \left( \{ \sigma^z_{ij} \} \right)} O \right) \label{rhoWM}
\end{equation}
\cxb{We also note that $\hat{\rho}$ can be expressed in a form compatible with what we have in the main text
\begin{equation}
\mathcal{E}_{\vect{x}}[\hrho_0] = (1 - p) \hrho_0 + p \left( \sum_m p_m \hat{P}_{m,\vect{x}} \hrho_0 \hat{P}_{m,\vect{x}} \right).
\end{equation}
if we make the identification
\begin{align*}
    &p = 2\cos t\sin t, \ \ \ p_+ = 1+2x, \ p_- = 1-2x\\
    &\hat{P}_{m,\vect{x}} = \frac{I+m \sigma^z_{ij}}{2}.
\end{align*}} Here $m = \pm 1$, and $\vect{x}$ corresponds to the link $<i,j>$ on the lattice. 

We can view $\tr(\hrho)$ as a partition function of a $3d$ classical Ising model from which one can extract $\tr(\hrho O)$. To see this, we note that the expectation of a string of Pauli-$\sigma^z_{ij}$ operators with respect to $\hrho$ is 1 when the string is closed and 0 when the string is open. \cx{This is because the wave function of the $3d$ toric code is a superposition of loop-configurations generated by the product of $B_p$, which only involves closed loops.} There are additional subtleties caused by logical operators, but there are only a finite number of them and will not affect the partition function in the thermal dynamic limit. Therefore, the partition function becomes
\begin{equation}
    \sum_{\mathcal{C}}(\tanh \tilde{\beta})^{\abs{\mathcal{C}}}
\end{equation}
where $\mathcal{C}$ denotes closed loops formed by links of $3d$ cubic lattice, and this partition function is exactly the high-temperature expansion of a $3d$ classical Ising model. 

\cx{Physically, the operator $\sigma^z_{ij}$ enables the ``hopping" of the charges of the $Z_2$ gauge theory, hence at larger $\tilde{\beta} > \tilde{\beta}_c$ (which can be tuned by choosing $x$ and $t$), the partition function of the $3d$ classical Ising model enters the ``low temperature ordered" phase, i.e. the condensate of the electric charge of the $Z_2$ gauge theory. It is important to note that this transition belongs to the $3d$ Ising universality class, which is consistent with the picture described in the main text that weak measurement drives the electric charges to condense on the temporal slab (which is a $3d$ space) $\tau = 0$ in the path-integral formalism (Fig.~\ref{fig:lattice}). }

\cx{In Ref.~\cite{su2024} it is also shown that measurement without post-selection can drive a phase transition in the 2nd Renyi entropy of the $Z_2$ gauge theory, which is consistent with the physical picture that weak-measurement can drive condensation of the bound state of gauge charges on the two temporal slabs in the path integral formalism (Fig.~\ref{fig:lattice}). }

\section{Weak-measurement on the lattice model and its field theory description}

In this section we explicitly formulate the weak-measurement driven phase transition discussed in the main text on a lattice model. We take the $3d$ cubic lattice as an example. The local operators consist of rotors $\theta_i$ defined on the sites, and $a_{ij}$ defined on the links of the lattice, along with their canonical conjugates $n_i$ and $E_{ij}$
\beqn
    &\comm{E_{ij}}{e^{\ii a_{i'j;}}} = \delta_{ij,i'j'}e^{\ii a_{ij}}, \ 
    &\comm{n_i}{e^{\ii \theta_{i'}}} = \delta_{i,i'}e^{\ii \theta_i}
\eeqn Note that the rotors $\theta_i$ and $a_{ij}$ are periodically define, $i.e.$ $\theta_i = \theta_i + 2\pi$, $a_{ij} = a_{ij} + 2\pi$, while their conjugates $n_i$ and $E_{ij}$ take discrete integer eigenvalues. 

\begin{figure}
    \centering
    \includegraphics[width = 0.6\textwidth]{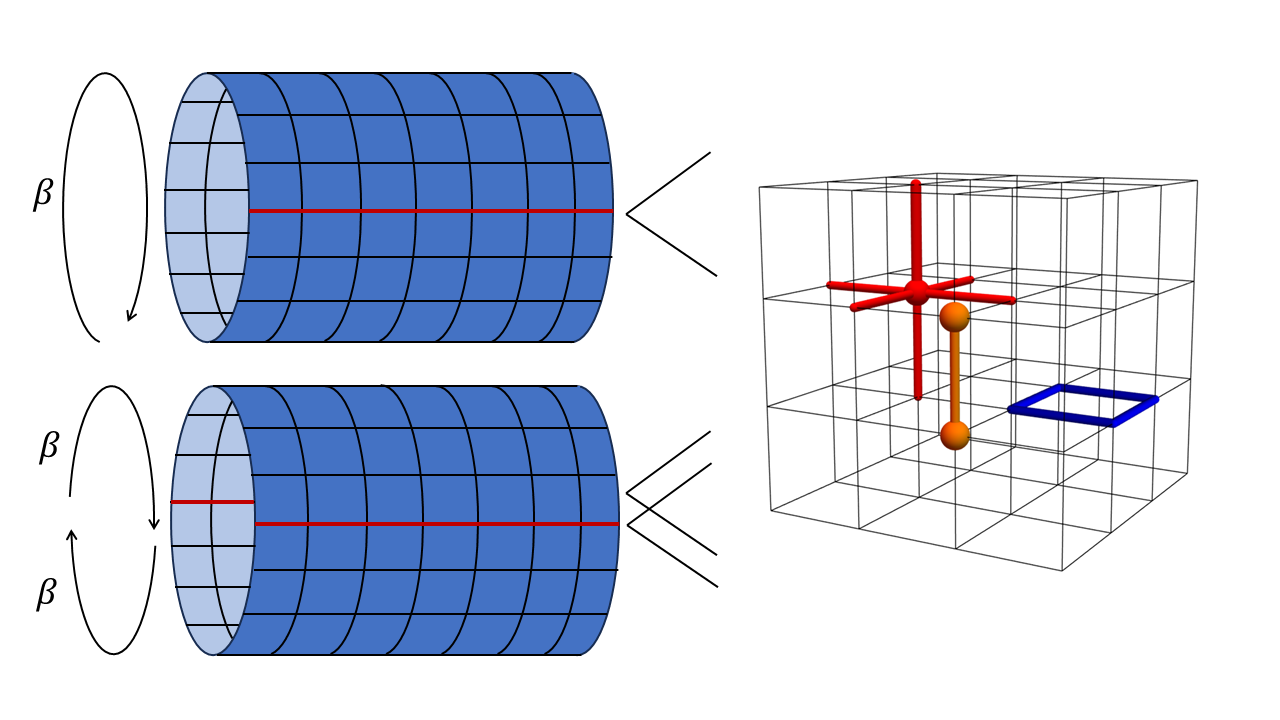}
    \caption{The upper left panel shows the protocol of our quantum operation with a biased selection process. This can be formulated as a defect insertion at a particular temporal slice in the Euclidean space-time path-integral (red line). The lower left panel shows the protocol that considers the 2nd R\'{e}nyi entropy under decoherence, which becomes defect insertion at two temporal slices separated by $\beta$ where $\beta\rightarrow\infty$. The right panel shows the spatial lattice structure. The red vertex stands for the Gauss law constraint operator, the blue plaquette stands for the plaquette term which controls pure gauge fluctuations, and the orange link is the coupling of the gauge field to the rotor matters.}
    \label{fig:lattice}
\end{figure}

Before weak measurement or decoherence, the pure quantum state $|\Psi\rangle$ we start with is the ground state of the following lattice Hamiltonian
\begin{equation}
    H = \frac{U}{2}\sum_i n_i^2 -J\sum_{i,\mu}\cos(\theta_i-\theta_{i+\mu} + a_{i,i+\mu}) + \frac{\tilde{U}}{2}\sum_{i,\mu}E_{i,i+\mu}^2 - K\sum_{i,\mu}\cos(\vect{\nabla} \times \vect{a}).
\end{equation}
The system is also subjected to a local ``Gauss law" constraint
\begin{equation}
    n_i - \vect{\nabla} \cdot \vect{E} = 0. \label{gauss}
\end{equation}
One can verify that the local operator $n_i - \vect{\nabla} \cdot \vect{E}$ commutes with the Hamiltonian. \cx{The phase diagram of the lattice gauge theory is well-understood: we choose that $U/J$ and $K/\tilde{U}$ to be greater than a critical value, so that the system is in its Coulomb or photon phase, i.e. 
Physically this means that the rotor charge is gapped, and the gauge field $\vect{a}$ is in its deconfined phase with gapless photon excitations. The structure of the spatial lattice is shown in fig.\ref{fig:lattice}. }

\subsection{Biased Selection}
We now describe a quantum channel that couples the system operator $\exp(\ii(\theta_i - a_{ij} - \theta_j))$ to an auxiliary qubit (one can consider more general ancilla degrees of freedom such as qudit or rotor but ancilla qubits suffice for our discussion). \cxb{The initial state is again prepared to be a tensor product between ground state of $H$ and $\ket{0}_a$ of all the ancilla qubits. }First, we have the qubit and $\theta_i, a_{ij}, \theta_{j}$ undergo a collective unitary of the following form (in what follows we denote $\Theta_{ij} = \theta_i - a_{ij} - \theta_{j}$). \\
\begin{equation}
    U = \frac{1}{\sqrt{2}}\begin{bmatrix}
        \cos t + e^{\ii \Theta_{ij}} \sin t , \ -\cos t + e^{\ii \Theta_{ij}} \sin t \\
        \cos t - e^{- \ii\Theta_{ij}} \sin t , \ \cos t + e^{-\ii \Theta_{ij}} \sin t 
    \end{bmatrix}
\end{equation}
One can verify this operator is unitary. When $t=0$, U takes the simple form
\begin{equation}
    U_0 = \frac{1}{\sqrt{2}}\begin{bmatrix}
        1 & -1 \\
        1 & 1
    \end{bmatrix}
\end{equation}
We can also rewrite $U$ in the following form
\begin{equation}
    U = e^{- \ii t H}\cdot U_0
\end{equation}
where \begin{equation}
    H = \ii e^{ \ii\Theta_{ij}}\sigma^{+}- \ii e^{- \ii\Theta_{ij}}\sigma^-.
\end{equation}

The density matrix we start with is
\begin{equation}
    \hrho_{\rm total} = \hrho_0 \otimes \ket{0}\bra{0}.
\end{equation}
Evolution with $U$ will generate an entangled state between the system and ancilla
\begin{equation}
    \frac{1}{\sqrt{2}}(\cos t + \sin t e^{\ii\Theta_{ij}})\ket{\psi}\ket{0} + \frac{1}{\sqrt{2}}(\cos t - \sin t e^{-\ii \Theta_{ij}})\ket{\psi}\ket{1}
\end{equation}
We then measure the ancilla in its computational basis which gives us an ensemble of states labeled by measurement outcome $s$, a string that records the measurement outcomes on each link
\begin{equation}
    \ket{\psi(s)} \propto \prod_{\rm link}\frac{1}{\sqrt{2}}(\cos t + s_{ij} \sin t e^{\ii s_{ij}\Theta_{ij}})\ket{\psi}
\end{equation}
To evaluate an average quantity of this ensemble, we need to investigate \begin{equation}
    \Bar{\expval{O}} = \sum_{s} P(s) \frac{\bra{\psi(s)}O \ket{\psi(s)}}{\braket{\psi(s)}{\psi(s)}}
\end{equation}
By Born's rule (no bias), this is equivalent to
\begin{equation}
    \tr \left( \prod_{\rm link} \frac{1}{\sqrt{2}}(\cos t + \sin t e^{\ii \Theta_{ij}})\hrho_0 \frac{1}{\sqrt{2}}(\cos t + \sin t e^{-\ii \Theta_{ij}})O + \prod_{\rm link} \frac{1}{\sqrt{2}}(\cos t - \sin t e^{- \ii \Theta_{ij}})\hrho_0 \frac{1}{\sqrt{2}}(\cos t - \sin t e^{\ii \Theta_{ij}})O \right)
\end{equation}
Assuming $O$ commute with $e^{\ii \Theta_{ij}}$, the equation above is equivalent to
\begin{equation}
    \prod_{\rm link} \tr (\hrho_0 O)
\end{equation}
hence the measurements without post-selection does not change the expectation value of $O$.

We now consider a biased selection, meaning we collect all measurement outcomes but we give a higher weight to the outcome with $s_{ij} = +1$. This is equivalent to
\begin{align*}
    &\tr \left( \prod_{\rm link} (\frac{1}{2}+x)(\cos t + \sin t e^{- \ii\Theta_{ij}})(\cos t + \sin t e^{ \ii\Theta_{ij}})\hrho_0 O + \prod_{\rm link} (\frac{1}{2}-x) (\cos t - \sin t e^{ \ii\Theta_{ij}})(\cos t - \sin t e^{- \ii\Theta_{ij}})\hrho_0 O \right) \\
    =&\tr \left( \prod_{\rm link} (1+ 4x \cos t \sin t \cos \Theta_{ij} )\hrho_0 O \right)
\end{align*}

We can again reorganize the mixed density matrix into the following form:
\begin{equation}
   \hrho \sim e^{- \frac{1}{2} I\left( \{ \Theta_{ij} \} \right) } \hrho_0   e^{- \frac{1}{2} I\left( \{ \Theta_{ij} \} \right) },  \ \ \  \tr\left( \hrho O \right) \sim \tr \left( \hrho_0 e^{- I \left( \{ \Theta_{ij} \} \right)} O \right) \label{rhoWM}
\end{equation}
\cx{In the path integral language this is equivalent to increasing the weight for the configurations with $\Theta_{ij} = 0$ on a $3d$ slice at $\tau = 0$ of the Euclidean space-time. By choosing $x$ and $t$, we can see that the weight of configurations with $\Theta_{ij} = 0$ can be arbitrarily larger than the weight of configurations with $\Theta_{ij} = \pi$, i.e. the measurement followed by selection can strongly prefer $\theta_i$ to order (charge condensation) at the temporal slab $\tau = 0$, similar to the previous section on the $Z_2$ gauge theory. }

\subsection{Field theory description}

For a (classical) XY model defined on the lattice, the partition funciton reas \beqn Z = \int_0^{2\pi} D\theta_i \exp\left( - I \left( \{ \Theta_{ij} \} \right) \right), \eeqn where $\Theta_{ij} = \theta_i - \theta_j$, and $I \left( \Theta_{ij} \right)$ is a periodic function of $\Theta_{ij}$. The simplest choice of $I( \{ \Theta_{ij} \} )$ is a nearest neighbor XY model: \beqn I(\{\Theta_{ij} \}) = \sum_{\rm n.n.} - (J/T) \cos(\theta_i - \theta_j). \eeqn On a $3d$ lattice, by tuning $J/T$ there is a phase transition between a condensate of $e^{\ii \theta_i}$ (for $J > J_c$), and a disordered phase with $J < J_c$. The $3d$ XY model is not soluble exactly, but its infrared physics can be studied reliably using field theory. In the field theory formalism, we need to use the ``coarse-grained" complex field $\phi(\vect{x})$ rather than the lattice operator $\exp(\ii \theta_i)$. The partition function is now mapped to a field theory \beqn Z \sim \int D \phi(\vect{x}) \exp\left( - \int d^d \vect{x} \ |\vect{\nabla} \phi|^2 + r |\phi|^2 + g |\phi|^4\right). \eeqn \cx{It is important to note that this field theory corresponds to a large class of lattice models with $\U(1)$ variables. For example one can add further neighbor interactions in the partition function, the change of the microscopic parameters will change the UV values of the parameters in the field theory, $e.g.$ $r$ and $g$. But the specific values of the field theory are not so important for the ``universal" physics in the infrared we are concerned about, because the physics in the infrared limit is controlled by the fixed point values of these parameters under renormalization group flow, regardless of the values in the microscopic (UV) scale. Hence the physics described by the fixed points of the field theory is a very generic description of a large class of models with $\U(1)$ variables.}  

If the original lattice model also involves gauge fields defined on the links, i.e. $\Theta_{ij} = \theta_i - \theta_j - a_{ij}$, with a lattice version of the Maxwell term of $a_{ij}$, the theory becomes an ``Abelian Higgs" model and has attracted a lot of interests in the past (see for example Ref.~\cite{abelianhiggs1,abelianhiggs2}). The field theory of this model is \beqn Z \sim \int D\phi(\vect{x}) D\vect{a}(\vect{x}) e^{- \cS[\vect{a},\phi]}, \ \ \ \cS[\vect{a},\phi] = \int d^dx \ |(\vect{\nabla} - \ii \vect{a}) \phi|^2 + r |\phi|^2 + g |\phi|^4 + \cL_{\rm Maxwell} \label{Sd}\eeqn

The pure state density matrix $\hrho_0$ of a quantum system before weak-measurement or decoherence can be represented in a standard path-integral form. If we are only interested in the physics in the infrared limit, the density matrix can be represented in a field theory form in the continuum, without loss of generality. 
In our case, the density matrix element between field configurations $| \vect{a}(\vect{x})_1, \phi(\vect{x}_1) \rangle $ and $| \vect{a}(\vect{x})_2, \phi(\vect{x})_2 \rangle$ reads
\beqn
&& \langle \vect{a}(\vect{x})_2, \phi(\vect{x})_2 |\hrho_0| \vect{a}(\vect{x})_1, \phi(\vect{x})_1 \rangle = \int D a_\mu(\vect{x}, \tau) D\phi(\vect{x}, \tau) e^{-S_{d+1}[a_\mu, \phi]}, \cr\cr && S_{d+1}[a_\mu, \phi] = \int d\tau d^dx \ |(\partial_\mu - \ii a_\mu)\phi|^2 + r' |\phi|^2 + g' |\phi|^4 + \frac{1}{4e^2} (f_{\mu\nu})^2,
\eeqn 
where $f_{\mu\nu} = \partial_\mu a_\nu - \partial_\nu a_\mu$, and we may introduce $a_0$ as a Lagrange multiplier to enforce the Gauss law constraint Eq.~\ref{gauss}. In the path-integral above, we also need the following boundary conditions 
$\vect{a}(\vect{x}, 0) = \vect{a}(\vect{x})_1$, $\vect{a}(\vect{x}, \beta) = \vect{a}(\vect{x})_2$, 
$\phi(\vect{x}, 0) = \phi(\vect{x})_1$, $\phi(\vect{x}, \beta) = \phi(\vect{x})_2$. In the bulk we need the system to be in the photon phase, hence we need $r' > 0$, i.e. the charged matter fields are gapped. For the physics in the infrared we can just focus on the last Maxwell term $\frac{1}{4e^2} (f_{\mu\nu})^2$ which leads to gapless photons in the bulk.

After weak-measurement described in the previous section, the evaluation of an operator reduces to the following path integral problem: \beqn \tr\left( \hrho O \right) \sim \int D\phi(\vect{x}, \tau) e^{- S[a_\mu, \phi] + \cS[\vect{a}, \phi]_{\tau = 0}} O_{\tau = 0}. \eeqn Here in the path integral we need to identify $\tau = 0$ and $\tau = \beta$, due to the trace. The extra term in the action $\cS[\vect{a}, \phi]_{\tau = 0}$ is a coarse-grained version of $e^{- I(\{ \Theta_{ij} \}) }$ in Eq.~\ref{rhoWM}, where $\Theta_{ij} = \theta_i - \theta_j - a_{ij}$ on the lattice, and the form of $\cS[\vect{a}, \phi]_{\tau = 0} $ is given by Eq.~\ref{Sd}. 

In the $d+1$ dimensional space-time bulk, the charges are gapped out, and the low energy physics is controlled by the Maxwell term only. But since the Maxwell term is a free theory, its effects on the temporal boundary can be understood reliably. This was discussed in (for example) Ref.~\onlinecite{sondual}, and the effect from the bulk will turn $\cS[\vect{a}, \phi]_{\tau = 0} $ in Eq.~\ref{Sd} to a nonlocal QED in Eq.4 of the main text. 


As we mentioned in the main text, weak-measurement with or without post-selection can also drive transition int the 2nd Renyi entropy, which is mapped to a path-integral with interactions between the two temporal slabs $\tau = 0, \beta$ (Fig.~\ref{fig:lattice}). In Ref.~\cite{wfdecohere,altman2,fan2023,su2024}, it was shown that weak-measurement can drive bound states of gauge charges on the two temporal slabs to condense. 
In our current problem, if we start with a state with an exact $\U(1)^{(1)}_m$ symmetry, weak measurements may potentially also drive a condensation of the bound state of electric charges $\phi$ on the $\tau = 0$ and $\tau = \beta$ slabs. This bound state condensation breaks the doubled IR-$\U(1)_e^{(1)}$ symmetry down to a diagonal IR-$\U(1)_e^{(1)}$ symmetry, using the language developed in Ref.~\onlinecite{sptdecohere,wfdecohere}. This transition is described by the following effective action: \beqn \cS &=& \int d^3x \ |(\vect{\nabla} - \ii \vect{a}_{1} - \ii \vect{a}_{2}) \phi|^2 + r |\phi|^2 + g |\phi|^4 \cr\cr &+& \frac{1}{2e^2} \vect{f}_{1} \cdot \frac{1}{|\partial|} \vect{f}_{1} + \frac{1}{2e^2} \vect{f}_{2}\cdot \frac{1}{|\partial|} \vect{f}_{2}. \eeqn $\vect{a}_{1}$ and $\vect{a}_{2}$ are the gauge fields on the $\tau = 0$ and $\tau = \beta$ interfaces respectively. $\phi$ is effectively coupled to one gauge field $\vect{a}_{+} = \vect{a}_{1} + \vect{a}_{2}$ with a nonlocal action, and the transition belongs to the same universality class as Eq.4 in the main text. The 2nd R\'{e}nyi entropy is the free energy for this transition, which should have a singularity at the transition.

In the condensate of $\phi$, $\vect{a}_{+} = \vect{a}_1 + \vect{a}_2$ acquires a mass and suppresses the correlation of field $\vect{B}_+ = \vect{B}_1 + \vect{B}_2$, i.e., a linear combination of fields on the two different temporal slabs. Hence we expect the following scaling: \beqn \log \left( \tr\{ \hrho \ e^{\ii B_a(\vect{0})} e^{-\ii B_b(\vect{x})} \hrho \ e^{ \ii B_a(\vect{0})} e^{-\ii B_b(\vect{x})} \} \right) \sim \frac{1}{|\vect{x}|^6}, \eeqn as this is the quantity that encodes the correlation of field $\vect{B}_+$. 


\section{Calculation with explicit form of density matrix}


In this section we compute quantities under weak measurement using the explicit form of density matrix. Let us start with the Hamiltonian of the quantized electromagnetic field in the $3d$ space: \beqn H = \int d^3x \ \frac{e^2}{2}
\vect{E}(\vect{x})^2 + \frac{1}{2e^2} \vect{B}(x)^2. \eeqn The
ground state wave functional of the EM field is \beqn |\Psi\rangle
\sim \int D \vect{a}(\vect{q}) \exp\left( -\cS[\vect{a}(\vect{q})] \right )| \vect{a}(\vect{q}) \rangle \text{ with } \cS[\vect{a}] = \frac{1}{2e^2} \int \frac{d^3q}{(2 \pi)^3}
 \frac{|\vect{q} \times
\vect{a}(\vect{q})|^2}{|\vect{q}|}. \label{GSwf} \eeqn
The correlation function between $\vect{B}$ fields in the ground
state reads \beqn \tr\{ \hrho_0
\vect{B}(\vect{0})\cdot\vect{B}(\vect{x}) \} \sim \int \frac{d^3q}{(2 \pi)^3} e^{i \vect{q} \cdot \vect{x}}  \epsilon^{aij} \epsilon^{amn} q_i q_m \langle a_j^T(\vect{q}) a_n^T(-\vect{q}) \rangle  \label{BBCorrelator}\eeqn
where $\langle a_j^T(\vect{q}) a_n^T(-\vect{q}) \rangle$ is the gauge-field propagator for the transverse modes of $\vect{a}$ computed according to the ``action" $2\cS [\vect{a}]$. It is straight-forward to check that $\langle a_j^T(\vect{q}) a_n^T(-\vect{q}) \rangle = \frac{e^2}{2} \frac{\delta^{jn} - q^j q^n/|\vect{q}|^2}{|\vect{q}|}$, hence
\beqn \tr\{ \hrho_0 \vect{B}(\vect{0}) \cdot \vect{B}(\vect{x}) \} \sim e^2 \int \frac{d^3q}{(2\pi)^3} e^{i \vect{q} \cdot \vect{x} } |\vect{q}| \sim \frac{e^2}{|\vect{x}|^4}. \eeqn


Suppose weak measurement drives electric charge to condense on the $(3+0)d$ temporal defect which effectively amounts to turning on a mass term for the gauge-field in the evaluation above, the correlation between magnetic fields reads as Eq.~\ref{BBCorrelator} but with the propagator of the transverse components of the gauge-field computed according to the modified ``action'' $2\cS[\vect{a}] + \frac{1}{2}m^2 \int \frac{d^3 q}{(2 \pi)^3} |\vect{a}(\vect{q})|^2$.
This new propagator -- in the presence of a gauge-field mass -- takes the form for the transverse components $\langle a_j^T(\vect{q}) a_n^T(-\vect{q}) \rangle = \frac{\delta^{jn} - q^j q^n/| \vect{q}|^2}{m^2 + 2 |\vect{q}|/e^2}$. Hence, at momentum much less than the scale set by the gauge-field mass
\beqn   \tr\{ \hrho
\vect{B}(\vect{0})\cdot\vect{B}(\vect{x}) \} &\sim&  \int \frac{d^3q}{(2\pi)^3} e^{i \vect{q} \cdot \vect{x}} \frac{|\vect{q}|^2}{m^2 + \frac{2}{e^2} |\vect{q}|} \cr\cr &\sim& \int \frac{d^3q}{(2\pi)^3} e^{i \vect{q} \cdot \vect{x}} |\vect{q}|^2  \left( \frac{1}{m^2} - \frac{2 |\vect{q}|}{e^2 m^4} + O(|\vect{q}|^2) \right). \eeqn Since the Fourier transformation of the quadratic piece $\vect{q}^2$ only leads to local correlations, the leading term that contributes at long distance is $O(|\vect{q}|^3)$ and yields \beqn \tr\{ \hrho \vect{B}(\vect{0})\cdot\vect{B}(\vect{x}) \} \sim \frac{1}{e^2 m^4} \frac{1}{| \vect{x}|^6}. \eeqn These evaluations are consistent with what we expected from the boundary criticality analysis. The evaluation of the electric field is similar, but we need to
treat $\vect{E}$ as $\delta/\delta \vect{a}(\vect{x})$ in the
evaluation with the wave functional. Then we shall see that the
extra mass term only causes some local correction to the
correlation between electric fields.

We now turn to the evaluation of the 't Hooft loop. We begin with the calculation in the pure state with $\U(1)^{(1)}_m$ SSB. We may define the eigenstates of the $\vect{E(x)}$ field, which should obey the following relation
\begin{equation}
    \braket{\vect{a}}{\vect{E}} = e^{-\frac{\ii}{2\pi}\int d^3x \vect{a}\cdot\vect{E}}
\end{equation}
The 't Hooft loop we would like to evaluate is
\begin{equation}
    \expval{e^{\ii\int_\cC d\vect{l}\cdot\vect{b}}} = \expval{e^{\ii\int_\cA d\vect{\sigma}\cdot\vect{E}}} = \int D \vect{E}(\vect{x}) \text{ } \frac{\braket{\Psi}{\vect{E}}e^{\ii\int_\cA d\vect{\sigma}\cdot\vect{E}}\braket{\vect{E}}{\Psi}}{\braket{\Psi}{\Psi}} \label{tHooft1},
\end{equation}
where $\vect{\sigma}$ is a surface element on membrane $\cA$. For simplicity, we choose $\cA$ to lie on the $z=0$ $XY$-plane. We can substitute in the ground-state wavefunction Eq.~\ref{GSwf} to get


\begin{align}
    \expval{e^{\ii \int d\vect{\sigma}\cdot\vect{E}}} = \frac{\int D\vect{a} D\vect{a'} D\vect{E} \text{ } e^{-\mathcal{S}[\vect{a}]-\mathcal{S}[\vect{a'}]-\frac{\ii}{2\pi}\int d^3x (\vect{a}-\vect{a'})\cdot\vect{E} + \ii \int d^3x \delta(z)\theta(\sqrt{x^2+y^2}<R)E_z}}{\int D\vect{a} D\vect{a'} D\vect{E} \text{ } e^{-\mathcal{S}[\vect{a}]-\mathcal{S}[\vect{a'}]-\frac{\ii}{2\pi}\int d^3x (\vect{a}-\vect{a'})\cdot \vect{E}}}
\end{align}
Integrating out field $\vect{E}$ in the numerator gives
\begin{equation}
    \vect{a}(\vect{x}) - \vect{a'}(\vect{x}) = 2\pi\delta(z)\theta(\sqrt{x^2+y^2}<R)\hat{e}_z = 2\pi T(\vect{x})\hat{e}_z \label{eq:T} = 2\pi \vect{T}(\vect{x})
\end{equation}
while the integration of $\vect{E}$ in the denominator simply imposes constraint $\vect{a}(\vect{x}) = \vect{a'}(x)$.
So the expectation value of the 't Hooft loop of Eq.~\Ref{tHooft1} reduces to
\begin{equation}
    \frac{\int D\vect{a} \text{ } e^{-\mathcal{S}[\vect{a}]-\mathcal{S}[\vect{a}+2\pi\vect{T}]}}{\int D\vect{a} \text{ }e^{-2\mathcal{S}[\vect{a}]}} \label{loop}
\end{equation}
with
\begin{equation}
    \mathcal{S}[\vect{a}] + \mathcal{S}[\vect{a}+2\pi\vect{T}] = 2\mathcal{S}[\vect{a}] + \frac{1}{2e^2}\int \frac{d^3q}{(2\pi)^3} \frac{(\vect{q}\times2\pi\vect{T(q)})\cdot (\vect{q}\times\vect{a(-q)})+\text{c.c.}}{|\vect{q}|} + \frac{2}{e^2}\int \frac{d^3q}{(2\pi)^3}\frac{\pi^2|\vect{q}\times \vect{T(q)}|^2}{|\vect{q}|}\label{eq:compact}
\end{equation}
After shifting the gauge-field in the path integral by $\vect{a}(\vect{q}) \rightarrow \vect{a}(\vect{q}) - \pi \vect{T}(\vect{q})$, we see that $\vect{a}(\vect{q})$ and $\vect{T}(\vect{q})$ decouple such that the expectation value is now merely
\begin{equation}
    \langle e^{\ii \int d \vect{\sigma} \cdot \vect{E} }\rangle =  e^{-\frac{1}{e^2}\int \frac{d^3q}{(2\pi)^3}\frac{\pi^2 (q_y^2+q_x^2) T(\vect{q})^2}{|\vect{q}|}}\label{eq:int1}
\end{equation}
The Fourier transform of $T$ is
\begin{align}
    T(q) = \int_0^R dr d \theta \text{ } re^{\ii q_2r\cos\theta} = \frac{2\pi R J_1(q_2R)}{q_2}
\end{align}
where $q_2 = \sqrt{q_x^2+q_y^2}$ is the magnitude of the in-plane component of the momentum, i.e. the Fourier distribution of $T$ is constant in $q_z$. $J_1(x)$ is the Bessel function of the first kind. We impose a cutoff $\Lambda$ for $|\vect{q}|$ so that the integral\ref{eq:int1} is convergent. The integral in the exponential can be simplified to
\begin{equation}
   \int d^3 q \text{ } \frac{q_2^2 T(\vect{q})^2}{|\vect{q}|} = \int d^3 q \text{ } (2 \pi R)^2 \frac{J_1^2(q_2 R)}{|\vect{q}|}
\end{equation}
A closed form expression is tedious, so we focus on the asymptotic behaviour of $J_1(q_2 R)$ which is 
\begin{equation}
    J_1(q_2 R)\sim \sqrt{\frac{2}{\pi q_2 R}}\cos(q_2 R - \frac{3\pi}{4}) + ...
\end{equation}
The large momentum integration would then roughly give
\begin{align*}
    \int dq_z q_2dq_2 \text{ } (2\pi R)^2 \frac{2}{\pi q_2 q R}
    \sim 8 \pi R\int  dq_zdq_2 \text{ }\frac{1}{q} \sim 16 \pi^2 R \Lambda
\end{align*}
One therefore get a perimeter law for the 't Hooft loop as expected:
\begin{equation}
    \expval{e^{\ii \int d\vect{\sigma}\cdot\vect{E}}} = e^{-\frac{2\pi \Lambda}{e^2}\mathcal{P}}.
\end{equation}

When the electric charges condense on the $(3+0)d$ temporal defect due to weak measurement, the evaluation of the 't Hooft loop follows the same path integral on the $(3+0)d$ as Eq.~\ref{loop}, but with a modified effective action \beqn \cS = \frac{1}{2e^2} \int \frac{d^3q}{(2 \pi)^3} \frac{|\vect{q} \times \vect{a}(\vect{q})|^2}{|\vect{q}|} - m^2  \int d^3x \sum_i \cos(\vect{a}_i) \eeqn
In the limit of strong $m^2$ (deep inside the Higgs phase) it suffices to take $\vect{a}(q)\sim \delta(q)$ (analogous to the derivation of Ref.~\onlinecite{altman1}), which would make the second term in eq.~\ref{eq:compact} vanish. This leaves us with the third term which differs from the massless case (eq.~\ref{eq:int1}) by a factor of 2. In other words the 't Hooft loop is now
\begin{equation}
    \expval{e^{\ii\int d\vect{\sigma}\cdot\vect{E}}} = e^{-\frac{4\pi \Lambda}{e^2}\mathcal{P}}.
\end{equation} Hence the 't Hooft loop still decays as a perimeter law as was expected from the argument in the main text, but with a larger factor, i.e. the electric charge condensation on the $(3+0)d$ temporal defect induced by weak measurement still suppresses the SSB of the $\U(1)^{(1)}_m$ symmetry.

\bibliography{SSBWM}

\end{document}